\documentclass[onecolumn]{aa}
\usepackage{graphicx}
\usepackage{txfonts}

\begin{document}

\title{A possible 3:2 orbital epicyclic resonance\\ 
in QPOs frequencies of Sgr A*}


   \author{
	Gabriel T{\"o}r{\"o}k\inst{1,2,3}
	}


   \institute{
        Nordita, Blegdamsvej 17, 
	DK-2100 Copenhagen, Denmark
        \and
        Department of Astrophysics, 
	G\"oteborg University, Sweden. \\(Address: Theoretical 
	Physics, Chalmers University, S-412-96 G\"oteborg, Sweden)        
        \and 
	Department of Physics,
        Bezru{\v c}ovo n\'am. 13, CZ-746 01 Opava, Czech Republic
	\\
	~ \email{terek@volny.cz}
	    }
\date{Received \dots; accepted \dots }

\abstract{ A recent measurement of double peak QPOs frequencies in Sgr A* is consistent with the 3:2 ratio. The same ratio is firmly established by previous observations in all double peak kHz QPOs in microquasars and theoretically explained by orbital epicyclic resonances excited in nearly Keplerian accretion flow in black hole's strong gravity. If confirmed, the 3:2 ratio of double peak QPOs in Sgr A$^*$ will be of a fundamental importance for the black hole accretion theory, by providing another clear argument that the accretion disk oscillations are indeed governed by non-linear, strong-gravity physics.}

\maketitle 

\section{Double peak QPOs with the 3:2 ratio in Sgr A* ?}

From the current analysis of stellar orbits within 10-100 light hours
of Sgr A$^*$, obtained independently by the MPI Garching group (\cite{Schoedeletal2002}, \cite{Schoedeletal2003}, \cite{Eisenhaueretal2003}) and the UCLA group (\cite{Ghezetal2003}, \cite{Ghezetal2004}) the best estimate of the black hole central mass is $3.6 \pm 0.4 \times 10^6~M_{\odot}$, where
the error bars represent both statistical and systematic errors. Earlier
lower statistical mass estimates based on proper motions of stars further
away gave somewhat lower masses ($2.6 \times 10^6 ~M_{\odot}$) but in light of new
information on stellar distribution and anisotropies these earlier data
would now also lead to masses near $3.5 \times 10^6 ~M_{\odot}$ (see the discussion in \cite{Schoedeletal2003}). This well constrained mass must be contained within a few light hours, i.e. several hundred Schwarzschild radii. The
analysis of the spatial distributiuon of the stellar cusp centered on the
BH suggests that most likely no more than $1\times 10^3\,M_\odot$ of that is in form of
stars or stellar remnants (the latter is less well constrained: \cite{Genzeletal2003}). From the lack of motion of the radio source itself and a theoretical
comparison of the stochastic motions of a BH of different masses with
surrounding stars, a lower limit of the mass contained within the radius
of the radio source (10 light minutes, 20 Schwarzschild radii) is about
$\sim 10^5 M_{\odot}$ (\cite{Reidetal1999}, \cite{Reidetal2003}, \cite{BackerandSramek1999}, \cite{Schoedeletal2003}). 

From these measurements and discussion, one concludes that the mass of the black hole in Sgr A$^*$ is most likely in the interval 

\begin{equation}
 2.6 \times 10^6 M_{\odot} < M < 4.4 \times 10^6 M_{\odot},
\label{equation1}
\end{equation}

\noindent and that a very conservative lower limit is $\sim 10^5 M_{\odot}$\footnote {We thank Reinhard Genzel for providing (in summer 2004) this updated discussion on the Sgr A$^*$ mass measurements.}.
\newline
\cite{Genzeletal2003} measured a clear periodicity of 17 min (1020 sec) in Sgr A* variability during a flaring event. This period is in the range of Keplerian orbital periods at a few gravitational radii away from a black hole with the mass constrained by (\ref{equation1}). More recently, \cite{Aschenbach2004a} have reported three other QPOs periodicities, 692 sec, 1130 sec, 2178 sec, roughly in the orbital Keplerian range, and two much shorter periods of 100 sec and 219 sec. The value of 1130 sec differ by 10\% from the 1020 sec period found by \cite{Genzeletal2003} and may correspond to the same periodicity of the source, but a firm conclusion is not possible because of the quality of data. With all reservation and caution that are necessary here, it was noticed (\cite{AbramowiczEtal2004a}, \cite{AbramowiczEtal2004c}; \cite{Aschenbach2004b}) that, 

\begin{equation}
(1/692):(1/110):(1/2178) \approx 3:2:1,
\end{equation}

\noindent i.e. that the ``Keplerian'' frequencies found in Sgr A* form ratios that are very close to be an exact commensurable sequence, 3:2:1. The  commensurability of QPOs frequencies in Sgr A*, if confirmed by a more accurate observations and data analysis, could be of a fundamental importance for a reason that we explain in the next section. 

\section{Commensurability of QPOs in microquasars: observations and theory}

\begin{figure*}[h]
\includegraphics[angle=-90, width=86mm]{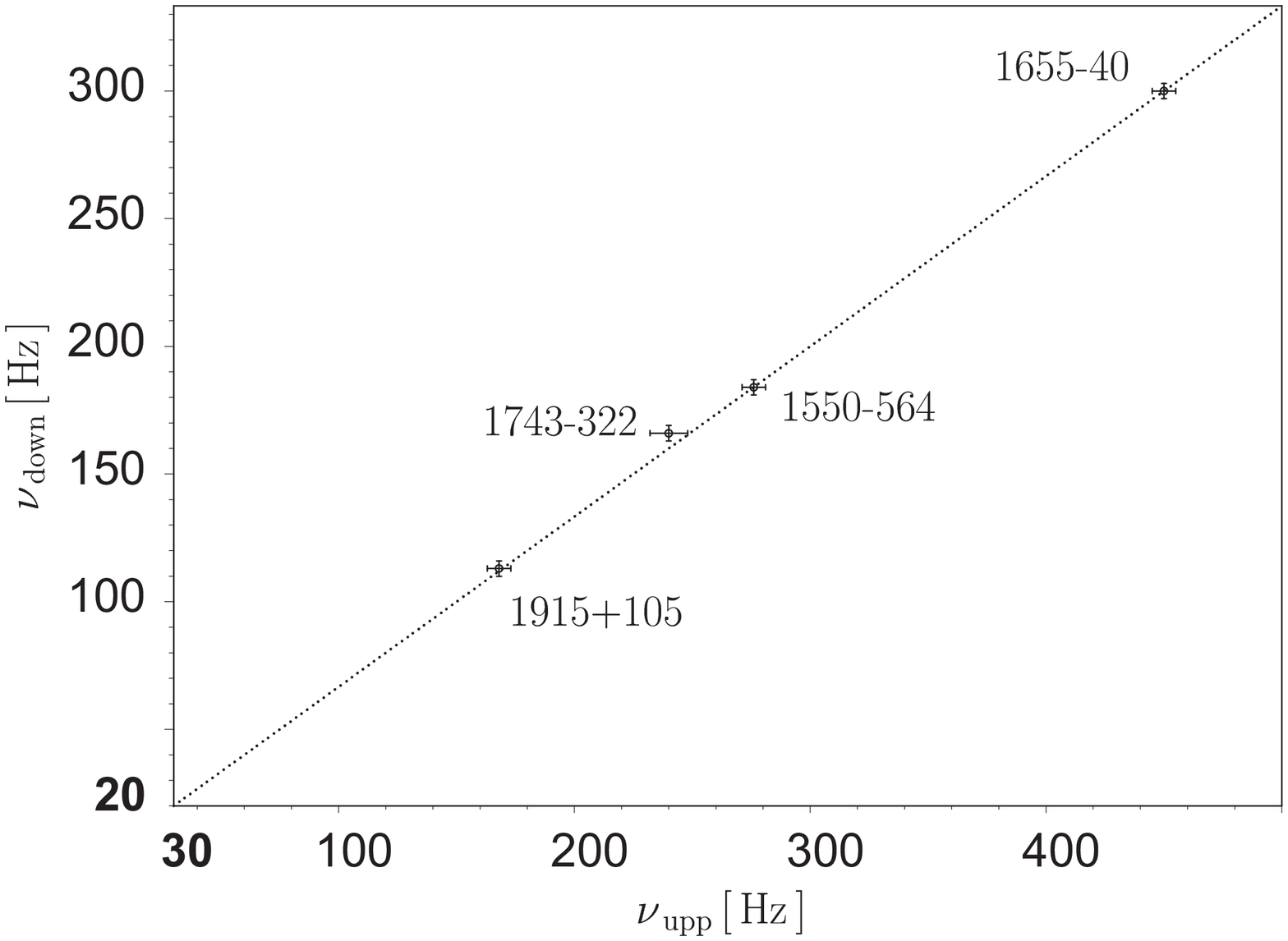}
\hfill
\includegraphics[angle=-90, width=86mm]{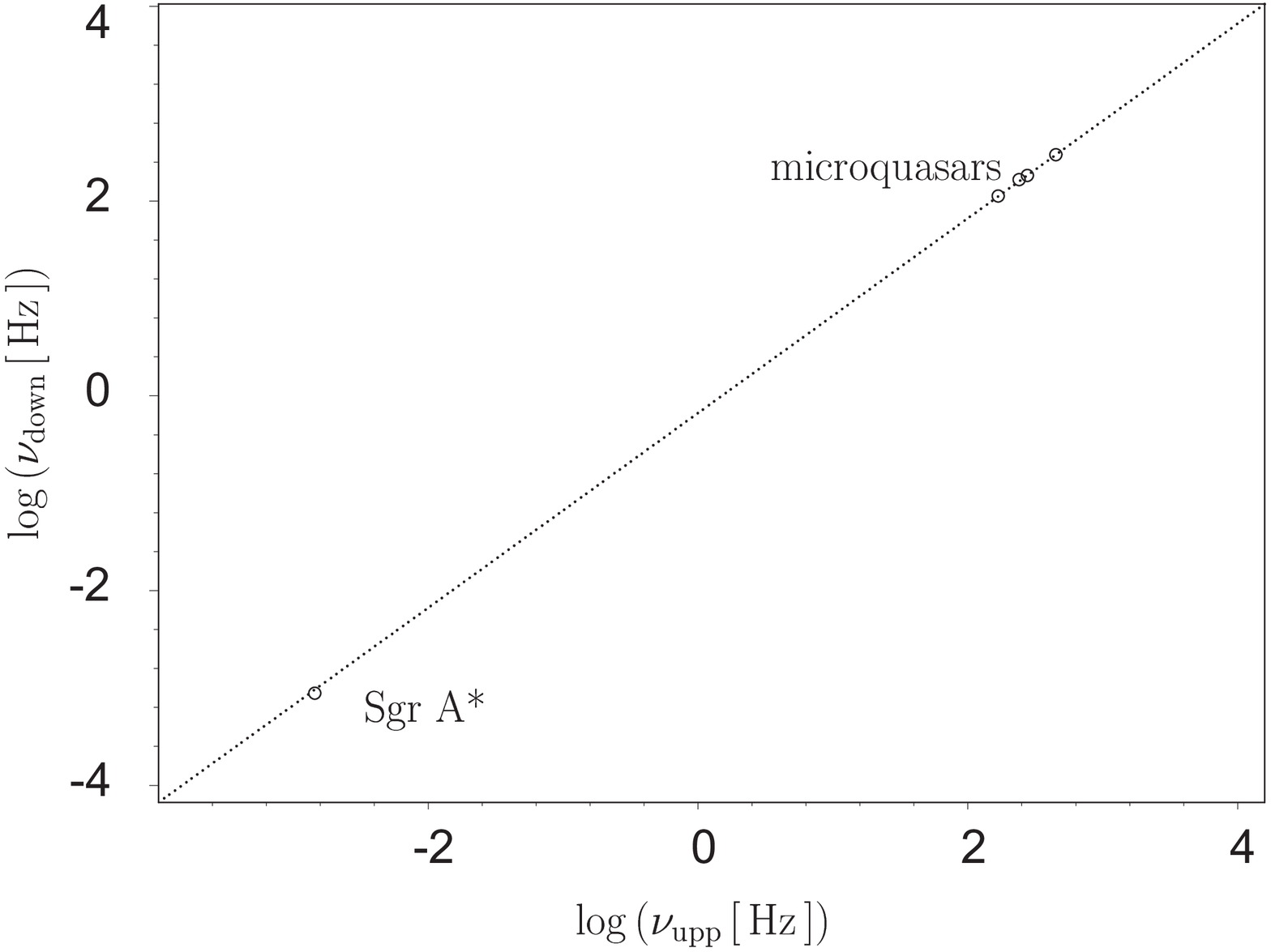}
\caption{\label{figure1}{\it Left:} In all four microquasars where double peak kHz QPOs were detected, the observed frequencies $\nu_{\rm upp}$ and $\nu_{\rm low}$ are clearly in 3:2 ratio.
{\it Right:} The same 3:2 ratio seems to be present in double peak QPOs in Sgr A*. 
The accuracy is so high that the error bars cannot be shown correctly in this logarithmic plot.
}
\end{figure*}

In the case of microquasars, the 3:2 ratio was found in all four sources with double peak QPOs detected. Table \ref{table1} and Figure  \ref{figure1} summarize the QPOs microquasars data relevant to the present Note. Impressively accurate ratio $\nu_{\rm upp}/\nu_{\rm down} = 3/2$ of frequencies was found in all four microquasars that display the double peak QPOs. In three microquasars with known mass $M$, the QPOs frequencies scale as (\cite{McClintockRemillard2003}),

\begin{equation}
\label{equation3}
\nu_{\rm upp} = 2.8 \left ({M \over M_{\odot}}\right)^{-1} {\rm [kHz]}.
\end{equation}

\begin{table}[h]
\begin{center}
      \caption[]{\label{table1}Frequencies of twin peak QPOs in microquasars and Galaxy centre black hole}
\renewcommand{\arraystretch}{1.2}
\begin{tabular}{| l || c  c | c  c | c | l |}
            \hline
    Source$^{~\rm (a)\,}$ &  $\nu_{\rm{upp}}\,$[Hz]&$\Delta\nu_{\mathrm{upp}}\,$[Hz]& $\nu_{\rm {down}}\,$[Hz]&$\Delta\nu_{\rm{down}}\,$[Hz]&$ {2\nu_{\rm{upp}}/3\nu_{\rm{down}}- 1}$& Mass$^{~\rm (b)\,}$ [\,M$_{\odot}$\,] \\
\hline
\hline
GRO~1655--40    & 450&$\pm\,3$& 300&$\pm\,5$ &\phantom{-}0.00000& $\phantom{1}$6.0\, --- $\phantom{1}$6.6   \\
XTE~1550--564   & 276&$\pm\,3$& 184&$\pm\,5$& \phantom{-}0.00000& $\phantom{1}$8.4\, --- 10.8   \\          
\phantom{RS}H~1743--322   & 240&$\pm\,3$& 166&$\pm\,8$&-0.03614& not measured    \\
GRS~1915+105    & 168&$\pm\,3$& 113&$\pm\,5$& \phantom{-}0.00885& 10.0 \,--- 18.0   \\
\hline
Sgr~A*\phantom{743--322}   & 1.445&$\pm 0.16$ mHz  &0.886&$\pm 0.04$ mHz  & \phantom{-}0.08728&2.6 --- 4.4\,$10^6$\\
            \hline
        \end{tabular}
\begin{list}{}{}
\item[$^{\rm{(a)}}$] Twin peak QPOs first reported by \cite{Strohmayer2001}, \cite{RemillardMunoMcClintockOrosz2002}, \cite{Homanetal2003}, \cite{RemillardMunoMcClintockOrosz2003}, and \cite{Aschenbach2004a}.
\item[$^{\rm{(b)}}$] See \cite{Greene2001}, \cite{Orosz2002}, \cite{Greiner2001},  \cite{McClintockRemillard2003}, and the first par of introduction.
\end{list}
\end{center}
\end{table}

Even before the double peak kHz QPOs have been discovered in microquasars (first by \cite{Strohmayer2001}), and the 3:2 ratio pointed out (first by \cite{AbramowiczKluzniak2001}), \cite{KluzniakAbramowicz2000} suggested on theoretical ground that these QPOs should have rational ratios, being due to resonances in oscillations of nearly Keplerian accretion disks. It seems that the resonance hypothesis is now well supported by observations, and that in particular the 3:2 ratio is seen most often in double peak QPOs in LMXB black hole and neutron sources, $2\nu_{\rm upp} = 3\nu_{\rm down}$.

According to the resonance hypothesis, the two modes in resonance have eigenfrequencies $\nu_{\rm rad}$, equal to the radial epicyclic frequency, and $\nu_{\rm v}$, equal to the vertical orbital frequency $\nu_{\rm{vert}}$ or the Keplerian frequency $\nu_{\rm K}$  (see \cite{KluzniakAbramowicz2004} for recent review). Several resonances of this kind are possible, and have been  discussed (see e.g. \cite{AbramowiczEtal2004b}). Main relations are summarized in Table \ref{table2}.

\begin{table}[h]
\begin{center}
      \caption[]{\label{table2}Relation for observed frequencies for standard~($\nu=\nu_\mathrm{vert}$) and ``Keplerian''~($\nu=\nu_\mathrm{K}$) resonances}
\renewcommand{\arraystretch}{1.2}
\begin{tabular}{| c | c | c | c || c | c |}
     \hline
    \multicolumn{4}{|c||}{Theory}
     & \multicolumn{2}{c|}{Observed frequencies}\\
     \cline{1-4}
    \multicolumn{2}{|c|}{Type of resonance} 
    &
     \multicolumn{2}{c||}
    {$\mathrm{n}\nu_{\mathrm{rad}}=\mathrm{m}\nu_\mathrm{}$}
     &\multicolumn{2}{c|}{}\\ 
     \cline{3-6}
      \multicolumn{2}{|c|}{}&~~~n~~~&~~~m~~~&
      $~~~\nu_\mathrm{upp}~~~~$ &$~~~\nu_\mathrm{down}~~~$ \\
     \hline
     \hline
     &parametric&3&2&$\nu_\mathrm{vert}$&$\nu_\mathrm{rad}$\\
     \cline{2-6}
   \rotatebox{90}{\makebox(0,0){~~standard}} 
   &3:1 forced&3&1&$\nu_\mathrm{vert}$&$\nu_\mathrm{vert}-\nu_\mathrm{rad}$\\
    \cline{2-6}
    &2:1
    forced&2&1&$\nu_\mathrm{vert}+\nu_\mathrm{rad}$&$\nu_\mathrm{vert}$\\
    \hline

     \hline
     &parametric&3&2&$\nu_\mathrm{K}$&$\nu_\mathrm{rad}$\\
     \cline{2-6}
   \rotatebox{90}{\makebox(0,0){~~Keplerian}} 
   &3:1 forced&3&1&$\nu_\mathrm{K}$&$\nu_\mathrm{K}-\nu_\mathrm{rad}$\\
    \cline{2-6}
    &2:1
    forced&2&1&$\nu_\mathrm{K}+\nu_\mathrm{rad}$&$\nu_\mathrm{K}$\\
    \hline
    \end{tabular} 
 \end{center}
\end{table}

\noindent Formulae for $\nu_{\rm vert}$ and $\nu_{\rm rad}$ in the gravitational field of a rotating Kerr black hole with the mass $M$ and spin $a$ are well known,

\begin{equation}
\label{frequencies}
\nu_{\rm vert}^2 = {\nu_{\rm K}^2}
\,\left(1-4\,a\,x^{-3/2}+3a^2\,x^{-2}\right),
~~~\nu_{\rm rad}^2 = {\nu_{\rm K}^2}\,\left(1-6\,x^{-1}+ 8 \,a \, x^{-3/2} -3 \, a^2 \, x^{-2} \right),
~~~\nu_{\mathrm{K}}={1\over 2\pi}\left ({{GM_0}\over {r_G^{~3}}}\right )^{1/2}\left( x^{3/2} + a \right)^{-1}.
\end{equation}

\noindent Here $x = r/(GM/c^2)$ is the dimensionless radius, expressed in terms of the gravitational radius of the black hole. 

For a particular resonance n:m, the equation $\mathrm{n}\nu_{\rm rad} = \mathrm{m}\nu_{\rm}$ ($\nu=\nu_\mathrm{vert}\,$or$\,\nu_\mathrm{K}$) determines the dimensionless resonance radius $x_{\mathrm{n}:\mathrm{m}}$ as a function of $a$.

\section{Application to Sgr A*}

\begin{figure*}[ht]
\includegraphics[angle=-90, width=85mm]{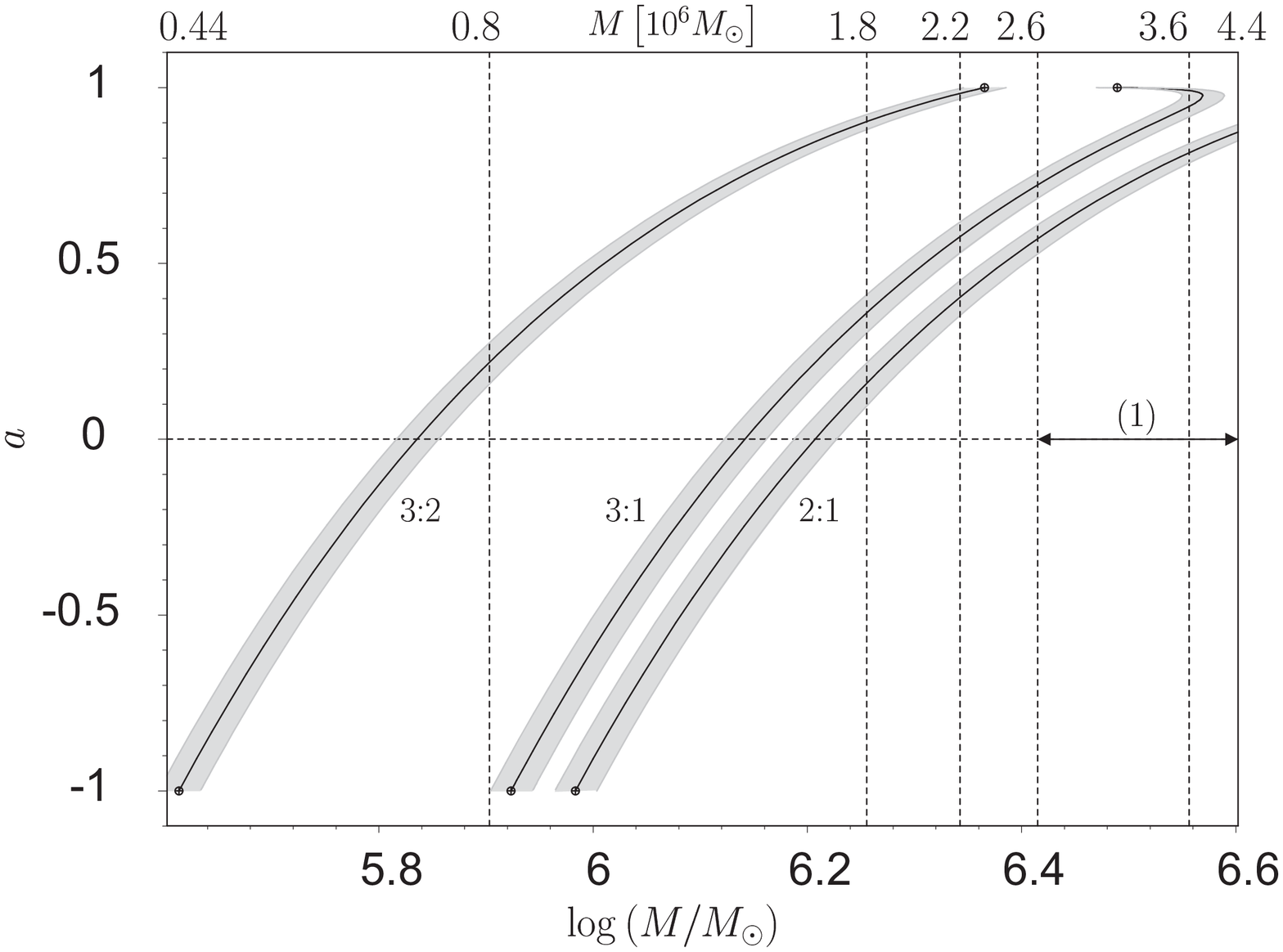}
\hfill
\includegraphics[angle=-90, width=93.8mm]{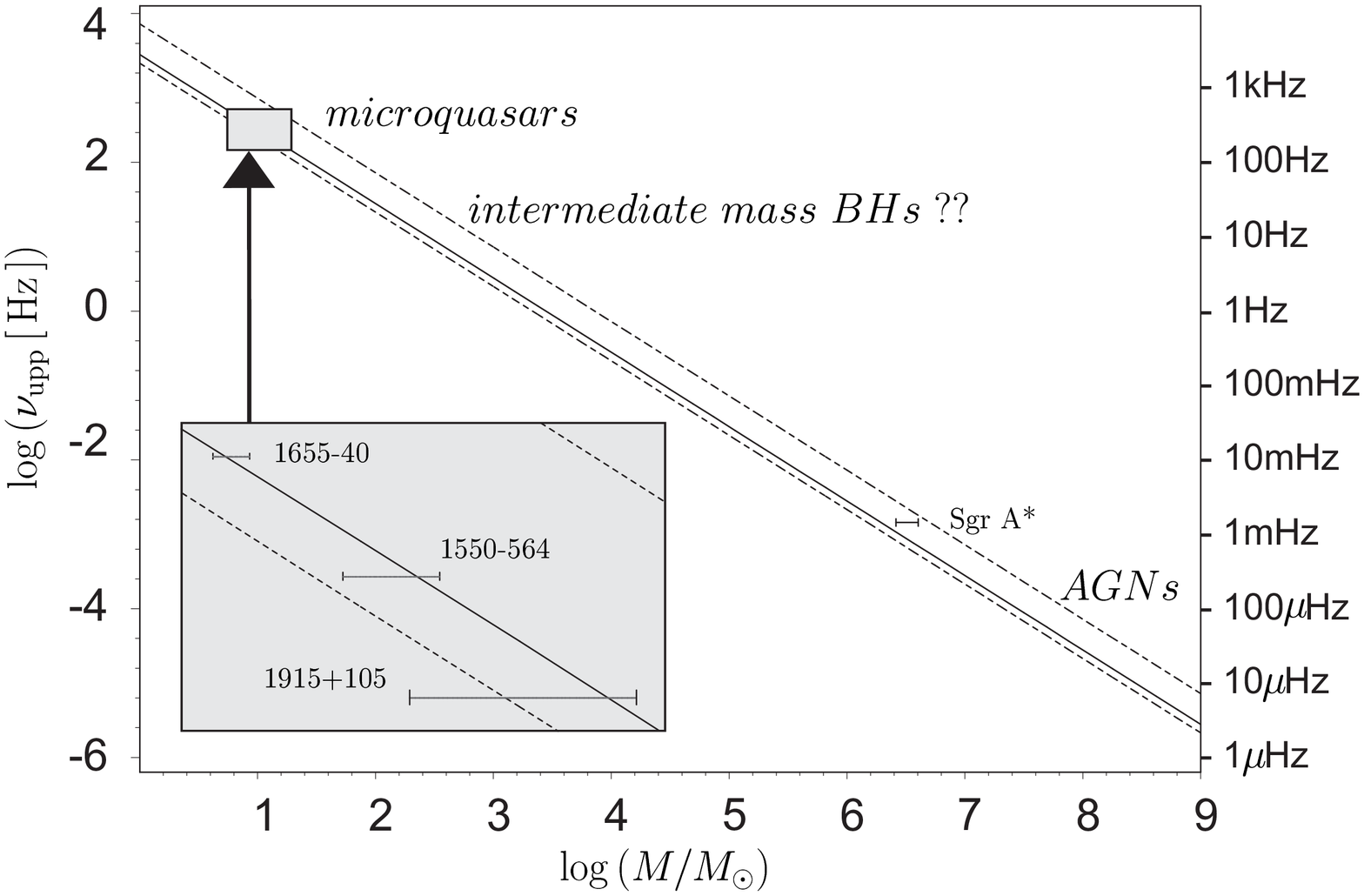}
\caption{\label{figure2}{\it Left:} Spin dependence for 3:2 parametric, 3:1 and 2:1 forced resonance in Sgr A* implied by measured $\nu_\mathrm{down}=0.886\,\mathrm{mHz}$, shadows respect accuracy of measuring.
{\it Right:} The case of the 2:1 parametric resonance for observed frequencies. Mass range for Sgr A* corresponds to the range given by inequality~(\ref{equation1}). Dotted lines are for $a=0\,$(lower line) and  $a=1\,$(upper line), the solid line is the best fit $\nu_{\rm upp}$ vs. $1/M$ found by \cite{McClintockRemillard2003} and described by equation~(\ref{equation3}).}
\end{figure*}

From the known mass of  Sgr A*, the observed $\nu_{\rm down}$ = $0.886 \,$mHz = $1/1130$~sec$^{-1}$, and from  the equation (\ref{frequencies}) on may calculate the black hole spin in SgrA$^*$, consistent with different types of resonances. This procedure was first applied to the microquasar GRO~1655--40 by \cite{AbramowiczKluzniak2001} and more recently for the other two microquasars by \cite{AbramowiczEtal2004b} and \cite{TorokEtal2005}. These results, together with these calculated in this Note for five representative values of the Sgr A* mass, are summarized in Table \ref{table3}, and illustrated in Figure \ref{figure2} for particular resonances.

{
\begin{table}[h]
\begin{center}
      \caption[]{\label{table3}Sgr A* spin estimates from observed 3:2 QPOs, calculated for five representative values of mass outgoing from large spectrum above lower conservative limit include the best mass estimate $3.6\times 10^6M_{\sun}$}
\renewcommand{\arraystretch}{1.2}
\begin{tabular}{| l || l | l | l | l | l | l | l |}
            \hline   
~&\multicolumn{7}{|c|}{$\mathrm{Mass ~[M_\odot]}$~:}
\\
\cline{2-8}
Resonance& $4.4\,10^5$& $0.8\,10^6$& $1.8\,10^6$& $2.2\,10^6$ & $2.6\,10^6$ & $3.6\,10^6$& $4.4\,10^6$ \\
\hline\hline
{3:2 [$\nu_{\theta},~\nu_r$] ~~ \phantom{``}parametric} & --- &0.22&0.90&0.98& --- & --- & --- \\  
\hline
{2:1 [$\nu_{\theta},~\nu_r$] ~~ \phantom{``}forced }& --- & --- &0.16&0.40&0.57&0.81&0.92\\ 
\hline
{3:1 [$\nu_{\theta},~\nu_r$] ~~ \phantom{``}forced }& --- & --- & 0.36 &0.58&0.72&0.95$\,$(0.99)$^*$& --- \\ 
\hline

\hline
{3:2 [$\nu_{\rm K},~\nu_r$] ~~``Keplerian'' p. }& --- &0.25& --- & --- & --- & --- & --- \\ 
\hline
{2:1 [$\nu_{\rm K},~\nu_r$] ~~``Keplerian'' f. }& --- & --- &0.16&0.41&0.58& 0.83&0.94\\ 
\hline
{3:1 [$\nu_{\rm K},~\nu_r$] ~~``Keplerian'' f. }& --- & --- &0.32&0.52&0.65&0.85&0.93\\          
     \hline
    \end{tabular}
\footnotesize{
\begin{list}{}{}
\item[*] see Figure \ref{figure2}.
\end{list}
}
\end{center}
\end{table}
}

\section{Discussion and conclusions}

If commensurability of double peak QPOs frequencies in Sgr A$^*$ is confirmed, this together with the already established $1/M$ scalling, would give a very strong support for the suggestion that the double peak QPOs physics, the same in microquasars and in Sgr A$^*$, is due to a non-linear orbital resonance in strong gravity. It would be interesting to see whether other black hole sources, ULXs and AGNs, show the same phenomenon (\cite{AbramowiczEtal2004a}).

For black hole sources with known mass that display the double peak QPOs, one may measure the black hole spin, but the spin estimate depends on which of the theoretically possible resonance, 2:1, 3:1, or 3:2, is actually excited in the source. At present, neither observations, nor the resonance theory could firmly determine this\footnote {\cite{Aschenbach2004b} argues that QPOs data suggests that black holes in three microquasars and Sgr A* listed in Table 1 have nearly the same spin $a \approx 0.99616$, due to a new relativistic effect that he found (a non monotonic behavious of orbital velocity with radius for rapidly rotating black holes, see also \cite{SlanyEtal2004}). Our calculations are based on standard types of non-linear orbital resonances, and do not include the Aschenbach effect.}.  

\begin{acknowledgements}

I thank Marek Abramowicz and Wlodek Kluzniak and also Zdenek Stuchlik and Vladimir Karas for discussion and help. The article was partialy written under the Erasmus-Socrates exchange program between Chalmers University and Silesian University at Opava. The final version was completed at Nordita.
\end{acknowledgements}

\end{document}